\documentclass[prl,twocolumn,superscriptaddress,showpacs]{revtex4}

\usepackage{graphicx}
\usepackage{dcolumn}
\usepackage{bm}

\usepackage{epsfig}

\begin{document}

\title{Creep and flow regimes of magnetic domain wall motion in ultrathin Pt/Co/Pt films with perpendicular anisotropy}

\author{P. J. Metaxas}
\email{metaxas@lps.u-psud.fr}
\affiliation{Laboratoire de Physique des Solides, Univ. Paris-Sud, CNRS, UMR 8502, F-91405 Orsay Cedex, France.
}%
\affiliation{
School of Physics, M013, University of Western Australia,
35 Stirling Hwy, Crawley WA 6009, Australia.
}%
\author{J. P. Jamet}%

\author{A. Mougin}%

\author{M. Cormier}%

\author{J. Ferr\'{e}}%

\affiliation{Laboratoire de Physique des Solides, Univ. Paris-Sud, CNRS, UMR 8502, F-91405 Orsay Cedex, France.
}%

\author{V. Baltz}
\author{B. Rodmacq}
\author{B. Dieny}
\affiliation{
SPINTEC, URA CNRS/CEA 2512,
CEA-Grenoble, 38054 Grenoble Cedex 9, France.
}%

\author{R. L. Stamps}
\affiliation{
School of Physics, M013, University of Western Australia,
35 Stirling Hwy, Crawley WA 6009, Australia.
}%

\date{\today}

\begin{abstract}
We report on magnetic domain wall velocity measurements in  ultrathin Pt/Co(0.5-0.8 nm)/Pt films with perpendicular anisotropy over a large  range of applied magnetic fields. The complete velocity-field characteristics are obtained, enabling an examination of the transition between thermally activated creep and viscous flow: motion regimes predicted from general theories for driven elastic interfaces in weakly disordered media. The dissipation limited flow regime is  found to be consistent with precessional domain wall motion, analysis of which yields values for the damping parameter, $\alpha$.
\end{abstract}

\pacs{62.20.Hg, 75.60.Ch, 75.60.Jk, 75.70.Ak}

\maketitle


Understanding the dynamics of an elastic interface driven by a force through a weakly disordered medium is a challenging
problem relevant to many physical systems. Examples include domain walls in ferromagnetic \cite{Lemerle1998,KrusinElbaum2001,Repain2004epl} and ferroelectric \cite{Paruch2006} materials, vortices in type-II
superconductors \cite{Blatter1994}, charge density waves \cite{Fukuyama1978} and contact lines during wetting of solids
by liquids \cite{deGennes1985}. While theory predicts three main regimes of motion \cite{Chauve2000,Blatter1994,Nattermann2001,Brazovskii2004} only the low force regime of creep has been experimentally studied through direct observation of the interface  \cite{Lemerle1998,KrusinElbaum2001,Paruch2006}. Regimes beyond that of creep, namely depinning and flow, have however been
evidenced  indirectly via ac susceptibility measurements \cite{Chen2002,Kleemann2007}. 
In this letter we  report on direct observation of magnetic domain wall motion  in ultrathin Pt/Co/Pt films over all motion regimes. This allows for a careful study of the wall velocity, in particular at the transition from creep to flow and in the  high field flow regime, where we consider the internal wall dynamics \cite{Sloncz1972,Schryer1974,bubblematerials}. 

\begin{figure}
\epsfig{file=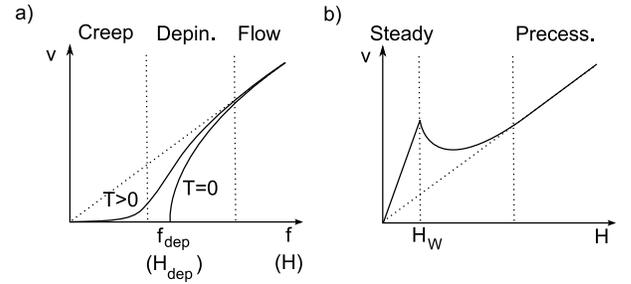,width=8cm}
\caption{\label{f:th} a) Theoretical variation of the velocity, $v$, of a 1D interface (domain wall) in a 2D weakly disordered medium submitted to a driving force, $f$ (magnetic field, $H$), at zero and finite temperature, $T$. The creep, depinning and flow regimes are labelled. b) Regimes of domain wall flow motion in an ideal ferromagnetic film without pinning. The steady and precessional linear flow regimes are separated by an intermediate regime which begins at the Walker field, $H_W$.}
\end{figure}

At zero temperature, an elastic interface in the presence of weak disorder will be pinned for all driving forces, $f$, below the depinning force, $f_{dep}$, at which a critical depinning transition \cite{Chauve2000} occurs  [Fig. \ref{f:th}a]. At finite temperature  the depinning transition becomes smeared due to thermal activation \cite{Brazovskii2004} and a finite velocity is then expected for all non-zero forces. This is true even for $f<<f_{dep}$ where the thermally activated interface motion is known as creep \cite{Chauve2000,Blatter1994}. At the other  extreme, once $f$ is sufficiently beyond $f_{dep}$, disorder  becomes irrelevant resulting in a dissipative viscous flow motion with $v \propto f$ \cite{Chauve2000}.  Ultrathin Pt/Co/Pt films with perpendicular anisotropy are systems in which one can easily study the field driven motion of quasi-1D domain walls (interfaces with elasticity due to their per-unit-length energy)  in a quasi-2D Ising system with appropriate weak quenched disorder due to nanoscale inhomogeneities \cite{Ferre2002,Lemerle1998,Kleemann2007}.


We have investigated domain wall dynamics in four such films with structure Pt(4.5 nm)/Co($t_{Co}$)/Pt(3.5 nm)  and Co layer thickness, $t_{Co}$, of 0.5, 0.6, 0.7 and 0.8 nm ($\pm0.05$ nm).  The films were sputter grown at $\sim$ 300 K on etched Si/SiO$_2$ substrates. Each film has a low density of efficient nucleation sites which allows us to measure domain wall motion at high field without excessive nucleation. The films' magnetic parameters are given in Table \ref{t}. First and second order effective perpendicular anisotropy fields for each
sample were determined using polar magneto-optic  Kerr effect  (PMOKE) anisotropy measurements \cite{grolier1993}. 
To estimate the wall width, $\Delta$, we  integrated these two fields into a total effective anisotropy field, $H_{eff}$, which includes the demagnetizing field of the perpendicularly saturated film. Out of plane PMOKE and SQUID hysteresis loops were used to determine each sample's coercive field, $H_C$, and saturation
magnetization, $M_S$, respectively. The Curie temperature of
each film, $T_C$, was deduced from the temperature dependence of the PMOKE signal at remanence.  On reducing $t_{Co}$, both $T_C$ and $M_S$ are also reduced compared to their bulk values of 1388  K and 1446 erg/G.cm$^3$
respectively \cite{MansuripurBook}, consistent  with the increasing 2D character of the films and the extent of   Co-Pt alloying. The exchange stiffness, $A$, was estimated from $T_C$ using a model for 2D  films with perpendicular anisotropy \cite{Bruno1992} and was found to be in quite good agreement with Brillouin light scattering experiments on thicker Co films \cite{Liu1996,Grimsditch1997}.


\begin{table*}
\begin{ruledtabular}
\begin{tabular}{ccccccccccccccccc}
\multicolumn{12}{c}{}&\multicolumn{2}{c}{Steady}&\multicolumn{2}{c}{Precessional}\\ \hline
t$_{Co}$&  $H_C$  & $H_{eff}$  & $K_{eff}$& $T_C$ & $M_S$ &$A$   & $\Delta$&$H_{max}$ & $H^*$ & $T_{dep}/T$ & $m$   &  $\alpha$ & $H_W$ & $\alpha$ & $H_W$ \\ \hline
 nm  & Oe &  kOe  & Merg/cm$^3$& K& erg/(G.cm$^3$) &  $\mu$erg/cm & nm  & Oe  &    Oe &  -  &m/(s.Oe)&- & Oe &   -  & Oe \\ \hline
0.5  &36 &  7.1 & 3.2& 415 & 910 & 1.4     & 6.2  & 1080   & 230     &  9  & 0.028 & 4.0 & 1690 & 0.27 & 120 \\
0.6   & 99 &   7.9 & 4.5&470 & 1130& 1.6   & 5.5 &  1670    & 590     &  14  & 0.026 & 3.7 & 2560 & 0.30 & 210 \\
0.7   & 195 &  5.3 & 3.2&520 & 1200 & 1.8   & 6.7 & 1930   & 750     &  22 & 0.034 & 3.5 & 2470 & 0.32 & 230 \\
0.8   & 280 &  3.1 & 2.0&570 & 1310 & 2.2   & 8.6 & 1420   & 650     &  35  & 0.043 & 3.5 & 2460 & 0.31 & 220 \\
\end{tabular}
\caption{\label{t}Co layer thickness, $t_{Co}$, coercive field, $H_C$ (field sweep rate during hysteresis loop measurement $\approx0.4$ kOe/s), total integrated effective perpendicular anisotropy field, $H_{eff}$, and the corresponding energy, $K_{eff}$,   Curie temperature, $T_C$, saturation magnetization, $M_S$ (error $\sim$ 10\%),  exchange stiffness, $A$, zero-field domain wall width, $\Delta=\sqrt{A/(K_{eff}+N_{y}2\pi M_S^2)}$ \cite{Mougin2007}, maximum possible applied field, $H_{max}$,  limit field, $H^*$,  for the validity of the creep velocity law (Eq. (\ref{eqn:vcreep})), the ratio of the depinning temperature, $T_{dep}=U_C/k_B$, to the experimental temperature, $T\sim 300$ K, high field wall mobility, $m$ (error $\sim$ 10\%) and  deduced values of the damping parameter, $\alpha$ (error $\sim$ 20\%), and Walker fields, $H_W$ (error $\sim$ 30\%), for the two possible types of  flow motion.}
\end{ruledtabular}
\end{table*}

\begin{figure*}
\includegraphics[height=4.2cm]{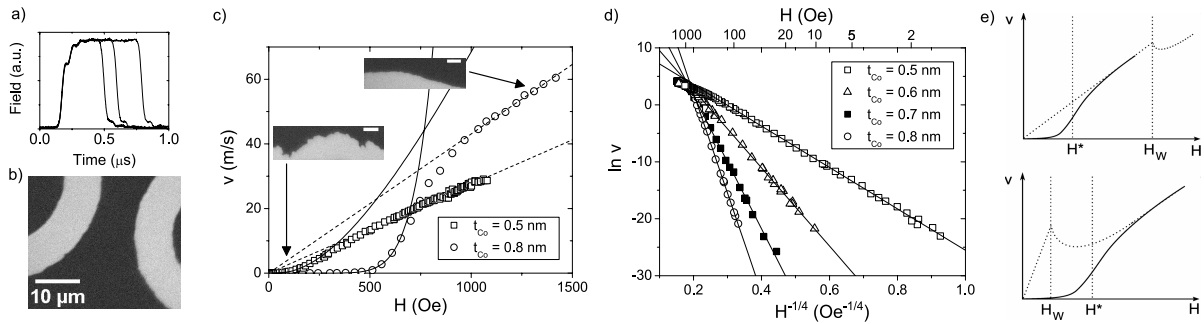}
\caption{\label{f:mainfig} a) Waveforms of field pulses of equal magnitude but differing duration used to move the walls. b) Subtracted domain image where the light area corresponds to the area swept out by the domain walls during a high field pulse. c) Domain wall velocity, $v$, versus applied magnetic field, $H$, for the $t_{Co}=0.5$ nm and $t_{Co}=0.8$ nm samples. We show a $v=mH$ fit to the high field flow data (dashed) and a fit of Eq. (\ref{eqn:vcreep}) to the low field creep data (solid line). The insets show domain images in the $t_{Co}=0.8$ nm film at  fields  which lie within the creep and flow regimes. The white scale bars are 5 $\mu$m long. d) Natural logarithm of the wall velocity versus the scaled applied field to demonstrate  low field creep. The solid lines are fits of the creep velocity expression  (Eq. (\ref{eqn:vcreep})). e) Examples of how the limited maximum applied field and  low field creep  prevent observation of multiple flow regimes (dotted curve) in the experimental data (solid curve - schematic only).}
\end{figure*}

To image the domains, we used a high resolution ($\sim$ 0.4 $\mu$m) far-field PMOKE microscope with a cooled CCD camera. We began by saturating the sample in a field of about -3 kOe. Reverse domains were  nucleated using positive high field  pulses ($\sim$ 1 kOe) generated using small coils mounted close to the sample surface. The domains were then expanded under the influence of additional positive  fields generated by either an electromagnet ($10^0-10^2$ Oe) or  by the pulse coils which can produce field pulses  up to $\sim2$ kOe with well defined plateaus and variable duration [Fig. \ref{f:mainfig}a].  The wall displacements were determined using a quasi-static technique in which an image of the domain structure was taken in zero field before and after the application of the field pulse. The domain structure was stable in zero field during imaging.  These images were then subtracted  from one another  [[Fig. \ref{f:mainfig}b] and the displacement measured.  We recorded wall displacements for successive field pulses of the same amplitude but increasing duration ($>250$ ns to ensure that the pulse waveform plateau was reached). The velocity was then determined from the gradient of a linear fit to the displacement data plotted against the pulse durations. In this way, effects from the transient parts of the pulses were eliminated. Fields were applied over times ranging from $\sim$ 250 ns to $\sim$ 10 h, chosen such that the displacement could be reliably determined ($>10$ $\mu$m) and the wall  remained  within the field of view (55 $\mu$m $\times$ 83 $\mu$m).  Above $H_{max}$ (Table \ref{t}) excess nucleation prevented reliable measurement of the wall displacement.


The  experimental  $v(H)$ curves for the $t_{Co}=0.5$ nm and $t_{Co}=0.8$ nm films are shown in Fig. \ref{f:mainfig}c. Velocity field characteristics are  qualitatively consistent with predictions (Fig. \ref{f:th}a): we evidence a low field, low velocity  regime and a high field, linear regime   separated by a smeared depinning region. 


We begin by verifying that the low velocity regime does in fact correspond to creep motion for which the following  velocity-field relationship is predicted  ($H<<H_{dep}$) \cite{Chauve2000}:
\begin{equation}
v =v_0 \exp\left[ -\left(\frac{T_{dep}}{T} \right) \left(\frac{H_{dep}}{H} \right)^{\mu} \right].
\label{eqn:vcreep}
\end{equation}
The depinning temperature, $T_{dep}$ is given by $U_C/k_B$ where $U_C$  is related to the height of the disorder-induced pinning energy barrier. $H_{dep}$ is the depinning field ($\equiv f_{dep}$), $\mu$ is a universal dynamic exponent equal to 1/4 for a 1D interface moving in a 2D weakly disordered medium \cite{Blatter1994,Chauve2000} and $v_0$ is a numerical prefactor \cite{Chauve2000}. To verify  the validity of  Eq. (\ref{eqn:vcreep}) with $\mu=1/4$ we plot  $\ln v$ versus $H^{-1/4}$  in Fig. \ref{f:mainfig}d.  
We indeed find linear behavior at low field  for all samples (corresponding to creep), proof of the universality of $\mu$, even upon varying  $t_{Co}$, $\Delta$, $H_{dep}$ and $T_{dep}$ (Table \ref{t}).


The extent of the linear region in Fig. \ref{f:mainfig}d is an interesting result. As can be clearly seen  in Fig. \ref{f:mainfig}c the creep law describes our data very well up to quite high fields, $H^*$ (given in Table \ref{t}), despite it being formally valid only in the small driving force limit ($H<<H_{dep}$) \cite{Chauve2000}.  $H_{dep}$ itself is difficult to determine at finite temperature because of the thermal smearing of the depinning transition. An additional problem  is that  $H_{dep}$ is related to magnetic anisotropy  \cite{Lemerle1998} which is itself  temperature dependent. As an estimate, we set $H_{dep}=H^*$ allowing us to calculate $T_{dep}/T$. We find that this ratio (probed in numerical simulations \cite{Kolton2005}) increases with $t_{Co}$ and that  $T_{dep}\geq9T$ (Table \ref{t}). 
It has been predicted that for $T_{dep}>>T$  an additional regime with $\ln v\propto H$ should exist in the vicinity of $H_{dep}$ \cite{Muller2001}.  However, we see no striking evidence of this  in our results. Note that such $\ln v\propto H$ behavior has been observed  in ultrathin Au/Co/Au systems in which pinning is stronger however the domain walls  did not exhibit creep motion \cite{Kirilyuk1993jmmm}.


An additional confirmation of the validity of interface theories to describe our results comes from an analysis of the wall roughness in the creep regime. This may be quantified by examining correlations between displacements of the wall from its mean position at points 
separated by a distance, $L$: $C^2(L)=\left\langle [u(x)-u(x+L)]^2 \right\rangle\propto(L/L_C)^{2\zeta}$ \cite{Huse11985,Huse21985}.  $u$ is the displacement of the wall from its mean position and $x$ is the co-ordinate along the direction of the wall's mean orientation. $L_C$ is a scaling length below which the interface is flat \cite{Lemerle1998}. The so-called wandering exponent, $\zeta$ \cite{Lemerle1998}, is predicted to have a value of $2/3$ for our system dimensionality \cite{Huse11985,Huse21985}. We have verified $C^2(L)\propto L^{2\zeta}$ for the $t_{Co}=0.6$ nm and $t_{Co}=0.8$ nm films and have found values for $\zeta$ of 0.7$\pm$0.1 and 0.66$\pm$0.06 respectively, in agreement with theory \cite{Huse11985,Huse21985} and previous experimental results \cite{Lemerle1998}.
In the creep regime the walls in the 0.5 nm film are smoother than those in the 0.8 nm film, testament to the 0.5 nm film's lower $T_{dep}$ and $H^{*}$.



In the high field regime, outside that of creep, walls in all of the films become significantly smoother [insets of Fig. 2(c)], indicative of the reduced relevance of the disorder. We find that the velocity in this high field regime can be fitted well using $v=mH$, with $m$ the wall mobility, consistent with predictions from moving interface theories. This is the first direct experimental measurement of interface flow in a weakly disordered system \footnote{Our flow regime corresponds to `slide' evidenced from ac susceptibility measurements in  \cite{Kleemann2007}.}.


Rather than disorder, dissipation limits the wall velocity in the flow regime. Here, dissipation is characterized by the magnetic damping parameter, $\alpha$, which is related to the wall mobility, $m$ \cite{Sloncz1972,Schryer1974,bubblematerials}.  In contrast to the standard elastic interface problem, in magnetic systems two separate regimes of linear flow are expected (each with $v=mH$ but different mobilites) [Fig. \ref{f:th}(b)].
This is due to an change in the internal dynamics of the wall above a critical field known as the Walker field, $H_W$ \cite{Sloncz1972,Schryer1974,bubblematerials}. Below $H_W$ the domain wall motion is steady with the mobility  given by $m=\gamma\Delta/\alpha$ where $\gamma$ is the gyromagnetic ratio  ($1.76\times10^{7}$ (Oe.s)$^{-1}$).  Sufficiently above $H_W$ there exists a second linear flow regime in which the magnetization within the domain wall precesses. The mobility in this precessional flow regime is lower than that of the steady one: $\left\langle m\right\rangle=\gamma\Delta / (\alpha+\alpha^{-1})$. These two linear flow regimes, together with a non-linear intermediate regime  \cite{Sloncz1972},  have  been recently observed in NiFe nanowires in which pinning is minimal \cite{Beach2005,Hayashi2007}. The Walker field can be written as $H_W=N_{y}2\pi\alpha M_S$ \cite{Mougin2007,Porter2004} where $N_{y}$ is the demagnetizing factor across the wall, given as  $t_{Co}/(t_{Co}+\Delta)$ in a simple approximation \cite{Mougin2007}.


The observation of only one linear regime in Fig. \ref{f:mainfig}c can be explained by two  scenarios. 
 Firstly, $H_W$ could be beyond $H_{max}$, meaning that the flow that we observe is steady flow (upper part of Fig. \ref{f:mainfig}e). From the steady flow mobility expression, we can then calculate $\alpha$, which we find to be about 4 for each film. The corresponding $H_W$ values are consistent with the steady flow assumption, being above $H_{max}$.
The other possibility is that $H_W$ is below $H^*$, in which case we would see  only  precessional flow  since the steady flow  would be obscured by the creep regime (lower part of Fig. \ref{f:mainfig}e). 
The precessional flow mobility expression yields $\alpha$ values on the order of 0.3, resulting in  $H_W$ values which are indeed below $H^*$. All values are given in Table \ref{t}.


On the basis of the relative magnitudes of the $\alpha$ values for each type of flow motion, we are  inclined to accept those corresponding to  precessional motion. Although  quite large, these $\alpha$ values are in fact on the same order of magnitude as those found  via other techniques in ultrathin Pt/Co/Pt multilayers \cite{Back1999JMMM,Barman2007} and thicker Cr-Pt-Co systems \cite{Back1999,Tudosa2004}. 
Very broad linewidths in ferromagnetic resonance (FMR) measurements prevented a reliable independent determination of $\alpha$ for the samples studied here. However, FMR measurements on slightly thicker Pt/Co/Pt films (with in-plane anisotropy) yielded  $\alpha$ values on the order of those in Table \ref{t} (eg. 0.22 for $t_{Co}=1.4$ nm).
  Considering the significant error in  $\alpha$ and limited range of $t_{Co}$ (due to a magnetic reorientation transition at $\sim$ 0.9 nm) we cannot comment seriously on  the dependence of $\alpha$ on $t_{Co}$ in Table \ref{t}. We also note   contributions from  less easily quantified errors in our determination of $A$, $N_{y}$ and $\Delta$.  Reduction of $H_{dep}$ could allow for direct  observation of the change in dynamics at $H_W$ which would confirm our  analysis and provide results complementary to recent measurements on in-plane magnetized nanowires \cite{Beach2005,Hayashi2007}.  Note  that our wall mobilities are smaller than those measured in NiFe nanowires \cite{Beach2005,Hayashi2007} and films \cite{Fukumoto2005} which may be attributed to the  narrower walls and higher damping in our films.


In conclusion, we have experimentally obtained for the first time the complete velocity-field characteristics of a 1D interface in a 2D weakly disordered medium through direct measurements of domain wall motion in  ultrathin Pt/Co/Pt films. This has allowed us to  examine both pinning and dissipation processes as well as test the validity of general theories concerning interface dynamics.  We also  observed  changes in the wall profile on moving from one motion regime to another. 
Finally, we determined a value for the magnetic damping parameter, $\alpha$, which describes the dissipation occurring during flow motion.

\begin{acknowledgments}
\small{
P.J.M. acknowledges  support from the Australian Government and a Marie Curie Action
(MEST-CT-2004-514307). P.J.M., A.M., J.F. and R.L.S. were supported by the FAST program (French-Australian Science and
Technology). P.J.M. and R.L.S. acknowledge the Australian Research Council. This work was partly done in the frame of the ACI contract, PARCOUR. We thank H. Hurdequint
for FMR measurements, R.C. Woodward for assistance with SQUID measurements, S. Wiebel for assistance with initial experiments and J.P. Adam  for useful discussions. }
\end{acknowledgments}


\end{document}